\documentclass[a4paper,fleqn,usenatbib]{mnras}

\usepackage[T1]{fontenc}
\usepackage{ae,aecompl}

\usepackage{graphicx}	
\usepackage{amsmath}	
\usepackage{amssymb}	

\title[Giant Broad Line Region in NGC 3998]{Photoionization Modelling of the Giant Broad-Line Region in NGC 3998}

\author[Devereux]{
Nick Devereux$^{1}$\thanks{E-mail: devereux@erau.edu}
\\
$^{1}$Embry-Riddle Aeronautical University, 3700 Willow Creek Road, Prescott, AZ 8301, USA\\
}

\date{Accepted 2017 September 22. Received 2017 August 20; in original form 2017 May 25}

\pubyear{2015}

\begin{document}
\label{firstpage}
\pagerange{\pageref{firstpage}--\pageref{lastpage}}
\maketitle

\begin{abstract}

Prior high angular resolution spectroscopic observations of the Low-ionization nuclear emission-line region (Liner) in NGC 3998 obtained with 
the Space Telescope Imaging Spectrograph (STIS) aboard the {\it Hubble Space Telescope (HST)}  revealed a rich UV--visible spectrum consisting of broad permitted and broad forbidden emission lines. The photoionization code XSTAR is employed together with reddening-insensitive emission line diagnostics
to constrain a dynamical model for the broad-line region (BLR) in NGC 3998. The BLR is modelled as a large H$^+$ region ${\sim}$ 7 pc in radius consisting of dust-free, low density ${\sim}$ 10$^4$ cm$^{-3}$, low metallicity ${\sim}$ 0.01 $Z/Z_\odot$ gas. Modelling the shape of the broad H${\alpha}$ emission line significantly discriminates between two independent measures of the black hole (BH) mass, 
favouring the estimate of \citet{Francesco2006}. Interpreting the broad H${\alpha}$ emission line in terms of a steady-state spherically symmetric inflow 
leads to a mass inflow rate of 1.4 ${\times}$ 10$^{-2}$ M${_{\sun}}$/yr, well within the present uncertainty
of calculations that attempt to explain the observed X-ray emission in terms of an advection-dominated accretion flow (ADAF). Collectively, the model provides an explanation for the shape of the H${\alpha}$ emission line, the relative intensities and luminosities for the H Balmer, [O\,{\sc iii}], and potentially several of the broad UV emission lines, as well as refining the initial conditions needed for future modelling of the ADAF.
 
\end{abstract}



\section{Introduction}

NGC 3998, the quintessential Liner \citep{Heckman1980}, is remarkable because high angular resolution spectra obtained with {\it HST} reveal not only broad permitted emission lines
but also similarly broad forbidden emission lines suggesting a gas density ${\sim}$10$^4$ cm$^{-3}$.
Such a low gas density allows the central UV--X-ray source to photoionize a large spherical volume of H, estimated to be ${\sim}$7 pc in radius \citep[][hereafter D11]{Devereux2011}. NGC 3998 is classified Liner 1.9 by \cite{Ho1997} due to the observation of a broad H${\alpha}$ emission line.
However, {\it HST}/STIS spectra presented in D11 reveal broad H${\beta}$ and H${\gamma}$ emission lines as well. Thus, a Liner 1.2 designation is more appropriate if the type scheme proposed by \cite{Whittle1992} for Seyferts is extended to include NGC 3998. 

Segregating various types of AGN spectroscopically involves ratios of forbidden and permitted emission line
fluxes \citep{Kewley2006, Baldwin1981}.  Such diagnostic diagrams are of limited utility
because ratios involving these lines depend on the angular size of the region being measured as discussed recently by \cite{Maragkoudakis2014}. Nevertheless, according to three diagnostic diagrams in \cite{Kewley2006}, the high angular resolution STIS spectra identify the AGN in NGC 3998 
ambiguously as an extremely unusual $\textrm{H\,\sc ii}$ type, 
a Liner, and on the borderline between Seyfert
and Liner. The ambiguity surrounding the true nature of the 
AGN in NGC 3998 motivates a more detailed examination.

The observation of broad Balmer emission lines has led several to propose that NGC 3998 possesses a bonafide broad line region (BLR) \citep{Heckman1980,Blackman1983a,Ho1997c,Francesco2006,Balmaverde2014}.
Neither the broad H Balmer lines nor the adjacent continuum are time-variable, at least on a timescale of  ${\sim}$ 4 years. NGC 3998
is just one of many non-reverberating AGN with an unusually large BLR size \citep{Devereux2015, Balmaverde2014,Zhang2007a, Wang2003c}. Non-reverberating AGN violate the correlation between BLR size and AGN luminosity established for reverberating AGN \citep{Kaspi2005},  posing a problem for all secondary methods of estimating BH masses that rely on a tight correlation between BLR-size and AGN luminosity \citep[e.g.][]{Vestergaard2006,Bentz2013, Ho2015}. The broad Balmer lines seen in non-reverberating AGN look very similar to those seen in reverberating AGN. Consequently, there is no easy way to distinguish them but using the BLR size-luminosity correlation established for reverberating AGN to estimate the BH mass in a non-reverberating AGN like NGC 3998 produces a wildly incorrect result as noted previously by D11. 

Because it is not possible to measure the size of the BLR in NGC 3998 using reverberation mapping, other methods have to be employed. One approach
demonstrated recently for the low luminosity Seyfert 1 nucleus in NGC 3516 \citep{Devereux2016} is based on photoionization modelling of a trifecta of key observations related to the H$^+$ gas responsible for producing the broad Balmer emission lines, namely the shape of the broad Balmer emission lines, the Balmer line luminosity, and the 
H${\alpha}$/H${\beta}$ ratio. Collectively, these three observables can be used to constrain a unique photoionization model for the H$^+$ gas. The principal merit of photoionization modelling is that it can be applied to all AGN regardless of whether they are reverberating or not. 
With a few exceptions \citep[][and references therein]{Pancoast2014a,Pancoast2014}, reverberation mapping generally yields only one measure of size, the Balmer reverberation radius \citep{Peterson2004}, but photoionization modelling yields two; a well defined inner and outer radius, along with a number of other interesting results concerning the physical
conditions of the H$^+$ gas. Given the importance of the BLR size-luminosity correlation, a key objective of this study is to employ the photoionization code XSTAR \citep{Kallman2001} to improve on the BLR size estimated previously (D11) for NGC 3998. A further objective is to explore whether or not photoionization modelling of the BLR is able to discriminate between the two independent measures of the BH mass in NGC 3998 that differ by almost a factor of four \citep{Walsh2012a, Francesco2006}. One may expect so because
the H$^+$ gas traces the velocity field to within ${\sim}$0.02 pc of the BH in NGC 3998, and the
outer radius of the H$^+$ region depends linearly on BH mass, everything else being equal.

The broader context of this study is to highlight the distinction between reverberating and non-reverberating AGN. One can use the lists of \cite{Maiolino1995, Ho1997c, Marziani2003} and \cite{Veron-Cetty2006} to infer
that reverberating AGN
represent less than half of all AGN with broad Balmer lines in a volume limited sample (D ${\leq}$ 40 Mpc). Thus, focussing on reverberating AGN and ignoring the larger population of non-reverberating AGN will inevitably lead to a biased perspective on BLR properties. In this regard, the non-reverberating AGN in NGC 3998 is spectacular because it appears to represent the very antithesis of the standard BLR paradigm articulated most recently by \cite{Schnorr-Mueller2016}. 

The layout of the paper is as follows. Section 2 introduces a new methodology that utilises reddening independent observables to constrain a unique XSTAR photoionization model for the BLR in NGC 3998, the physical properties of which are discussed further in Section 3.
Conclusions follow in Section 4.

\section{Photoionization Modelling}

The photoionization model described in the following for NGC 3998 features gas of low density (${\sim}$10$^4$ cm$^{-3}$) and high ionization parameter, very different from other photoionization models in the published literature that attempt to explain the broad emission lines of AGN in terms of high density (${\ge}$ 10$^9$ cm$^{-3}$), low ionization parameter gas that is contained in a broad-line ``cloud" or an ensemble of such ``clouds" \citep{Baldwin1995, Korista2000, Korista2004}. Those models have several shortcomings, the most persistent of which are the origin and confinement of such clouds. They also 
have difficulty explaining the high electron temperature (${\sim}$ 20,000 K) of photoionized gas in low ionization AGN \citep{Komossa1997, Ferguson1997, Richardson2014}. If such ``clouds" exist, in all likelihood they would not be in a vacuum but rather surrounded by a lower density interstellar medium that could be easily photoionized by the central UV--X-ray source. Thus, the geometry envisaged here for the BLR in NGC 3998 is akin to an $\textrm{H\,\sc ii}$ region or a planetary nebula, but with a much higher ionization parameter. 
In the following, XSTAR is used to model photoionization of the H$^+$ region anticipated to be responsible for producing the broad Balmer lines seen in NGC 3998. 

XSTAR is a computer code designed to model the emergent spectrum of an H$^+$ region that is photoionized by a centrally located UV--X-ray source. 
XSTAR parameters are listed in Table 1, a full description which can be found in the XSTAR documentation\footnotemark
\footnotetext{https://heasarc.gsfc.nasa.gov/xstar/docs/html/xstarmanual.html}. Values are listed in Table 1 only for fixed quantities and initial conditions.
Briefly, the parameters describe the shape and amplitude of the ionizing source, and the physical properties of the medium to be ionized.

The model involves a spherically symmetric distribution of neutral H gas that is photoionized by a centrally located UV--X-ray source. Spherical geometry is suggested by the observation (D11) that the observed Balmer line luminosity is comparable to that expected given the number of ionizing photons available from the central UV--X-ray 
source implying a high covering factor ${\sim}$1 for the BLR gas. A spherical geometry, as opposed to that of a thin disc, is further supported by the observation of single-peak broad Balmer emission lines (D11).  The radial number density for such a spherical distribution of neutral gas ${\rho}(r)$ is represented by a power law of index $n$, normalized by a number density ${\rho}_o$, at a reference radius $r_o$, so that ${\rho}(r)$=${\rho}_o(r/r_o)^{-n}$, where ${r}$ is the radial distance from the central UV--X-ray source.
The inner radius of the spherical model is determined by the initial choice of ionization parameter ${\xi}$ \citep{Tarter1969}, 
\begin{equation}
\xi = L/(\rho r^2)
\end{equation}
where $L$ is 
the ionizing luminosity. The ionization parameter used within XSTAR is different than the parameter $U$ employed, for example, by \cite{Ferland1983} 
\begin{equation}
U = Q(H)/(4 \pi \rho r^2 c )
\end{equation}
where $Q(H)$ is the number of H ionizing photons s$^{-1}$, and c is the speed of light. Converting between ${\xi}$ and $U$ depends on the frequency dependence of the ionizing continuum L$_{\nu}$  since
\begin{equation}
 Q(H) =  \int^{\nu_{max}}_{\nu_o} [L_{\nu}/ h \nu ]d\nu
\end{equation}
On the one hand the continuum emission of NGC 3998 has been interpreted in terms of an ADAF, \citep{Narayan1994} plus jets \citep{Nemmen2014, Younes2012a, Ptak2004a}. On the other hand, \cite{Maoz2007} and \cite{Pian2010} present evidence that the intrinsic visible--UV continuum of Liners may be similar to Seyferts, implicating geometrically thin accretion discs. 
\cite{Yu2011} demonstrate that the various continuum models actually look very similar to each other and present theoretical arguments in favor of an ADAF origin.  Consequently, the ADAF continuum of \citet{Nemmen2014}, illustrated in Figure 1, is adopted to represent the intrinsic UV--X-ray continuum of NGC 3998, the luminosity of which corresponds to 9 ${\times}$ 10$^{41}$ erg s$^{-1}$ integrated between 1 and 1000 Ryd yielding 7.53 ${\times}$ 10$^{51}$ H ionizing photons s$^{-1}$.
Thus, ${\xi}$ = 45 $U$.

Neither the gas pressure nor the gas density are held constant for any of the XSTAR models, in contrast to photoionization models of broad-line clouds.  Radiation pressure is negligible since the AGN in NGC 3998 radiates at 4 x 10${^{-4}}$ of the Eddington luminosity limit.  An upper limit to the integrated column density of 10$^{24}$ cm$^{-2}$ was imposed but never achieved for any
of the XSTAR models. 
The XSTAR code runs significantly faster by limiting the suite of elements to those for which emission lines are seen with high confidence in the STIS spectra;
H, He, C, N, O, Mg, and S. Also Fe which ${\it may}$ be present. Including all 30 elements available 
within XSTAR changes the model emission line luminosities by ${\sim}$ 0.04\% which is inconsequential compared to the measurement uncertainties.

\begin{figure}
	\includegraphics[width=\columnwidth]{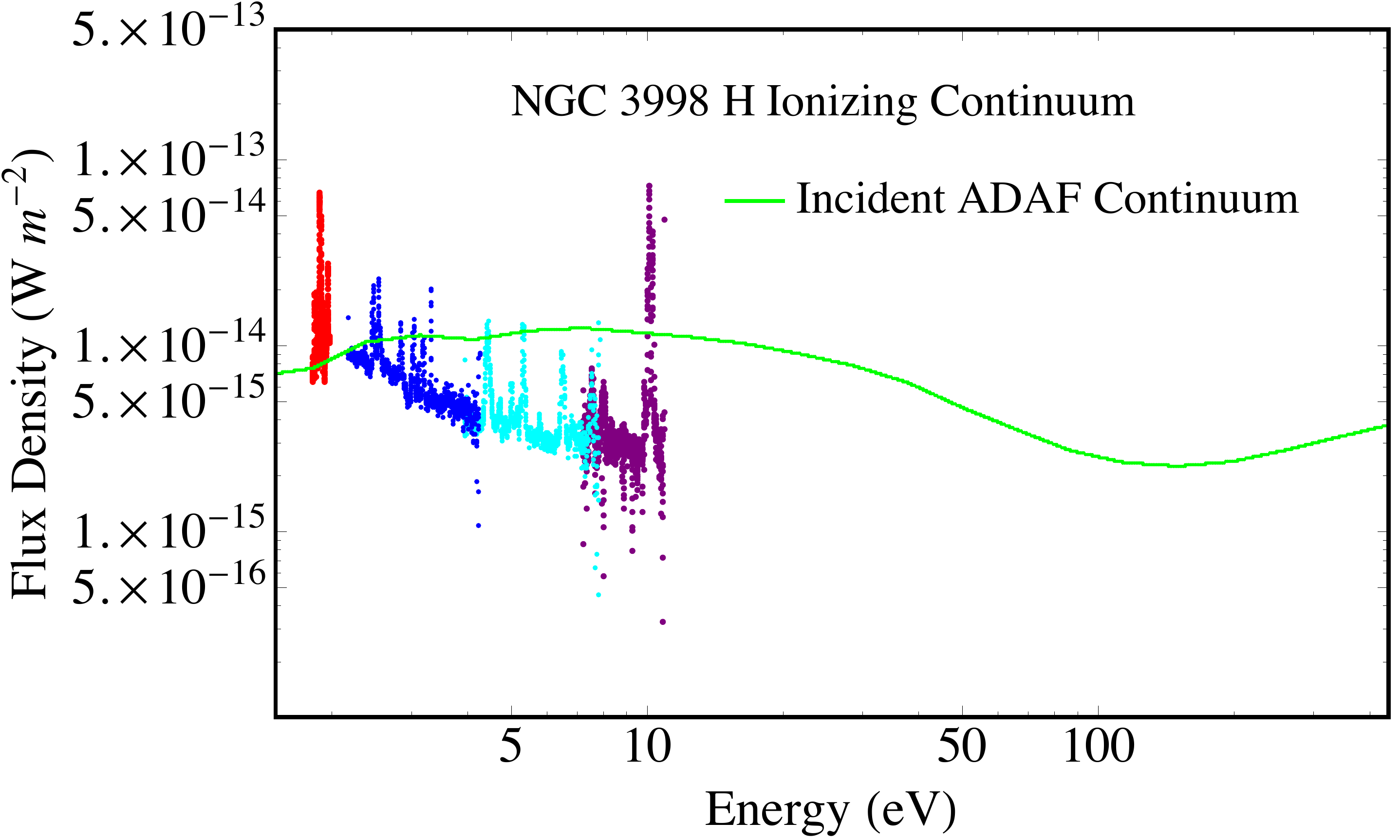}
    \caption{Observed UV--visible continuum of NGC 3998 defined by contemporaneous STIS observations with the G750M, G430L, G230L and G140L gratings, depicted by red, blue, indigo and purple dots, respectively. The solid green line represents the incident ADAF continuum of \citet{Nemmen2014}. Units of the ordinate are W/m$^2$ and the abscissa is eV.}
    \label{fig:example_figure}
\end{figure}

\begin{table}
\centering
\caption{XSTAR Input Parameters}
\begin{tabular}{ll}
\hline
Parameter & Value \\
\hline
covering fraction & 1 \\
temperature (10$^4$ K) & ... \\
constant pressure switch (1=yes, 0=no) & 0 \\
pressure (dyne cm$^{-2}$) & ... \\
density (cm$^{-3}$) & ... \\
spectrum type & file \\
spectrum file & ... \\
specrum units? (0=energy, 1=photons) & 0 \\
radiation temperature or alpha & ...  \\
luminosity (10$^{38}$) erg/s & 9e3 \\
column density (atoms cm$^{-2}$) & 1e24 \\
log (ionization parameter) & ... \\
density distribution power law index  & ... \\
\hline
\end{tabular}
\end{table}

XSTAR produces a comprehensive output including integrated emission line luminosities as well the radial dependence for a number of important physical quantities including the ionization fraction,
electron temperature, electron density and ionization parameter. 
Additionally, each model produced a unique Balmer line emissivity which was used to generate
a model H${\alpha}$ emission line profile
for comparison with a normalized version of the observed one. Details of the H${\alpha}$ emission line
profile modelling are provided in the next section.

\subsection{H${\alpha}$ Emission Line Profile}

NGC 3998 was visited twice with STIS and the
H${\alpha}$ emission line was observed with the G750M grating on both occasions. 
Here these two spectra have been combined in order to improve the signal-to-noise ratio.
The result reveals a distinctly triangular profile shape illustrated in Figure 2. 
The H${\alpha}$ emission line is the brightest and 
best resolved of all the Balmer lines, consequently it is the best line to model.

\begin{figure}
	\includegraphics[width=\columnwidth]{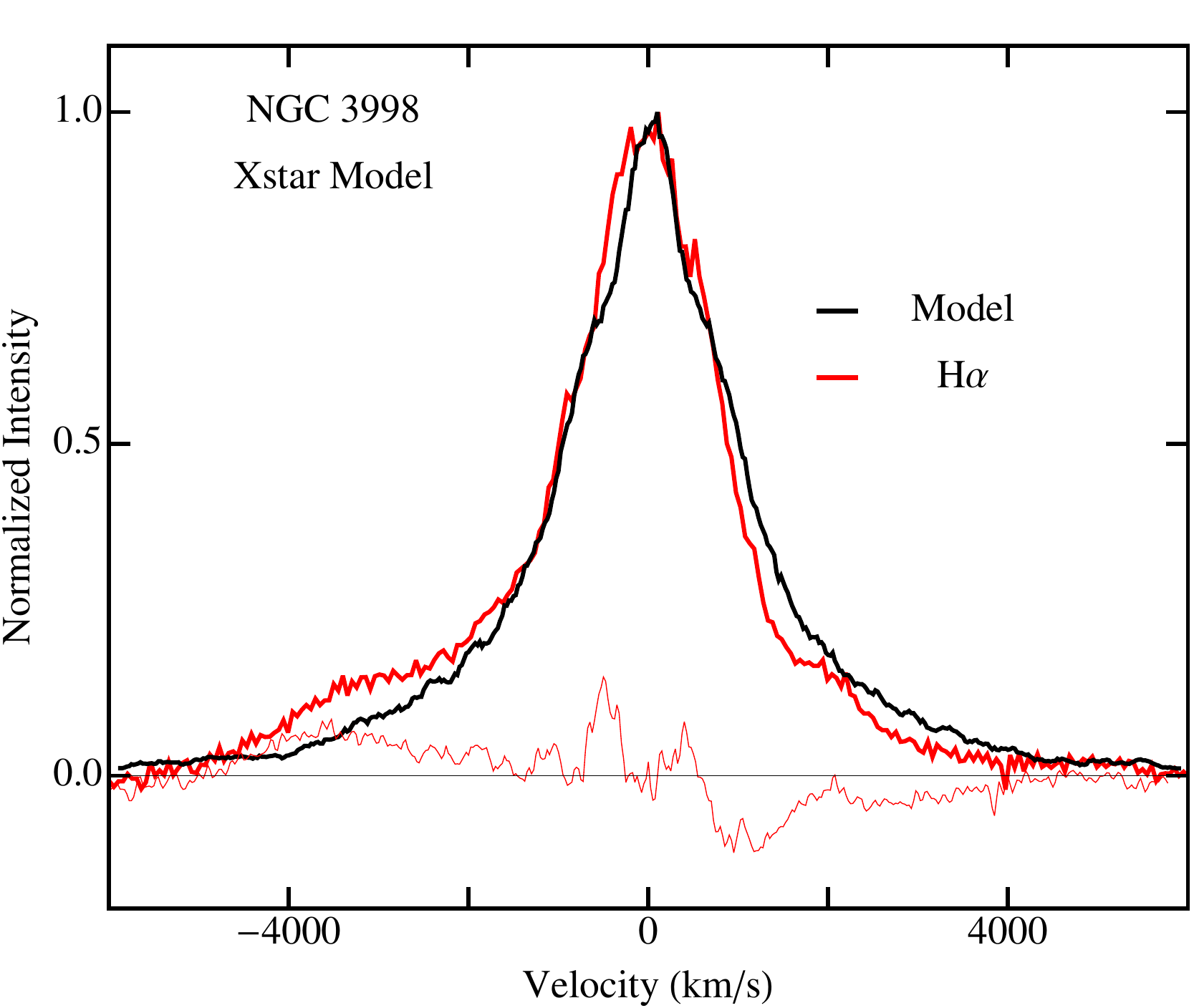}
    \caption{The observed normalized H${\alpha}$ emission line profile is depicted by the red line whereas
the XSTAR model is shown in black for ${\rho}_o$ = 3.55 ${\times}$ 10$^{4}$ cm$^{-3}$, $r_o$ = 8 ${\times}$ 10$^{-2}$ pc. The residual between the observed and
model line is represented by the thinner red line. The abscissa indicates radial velocity relative to rest in km/s (see Section 2.1 for details).}
    \label{fig:example_figure}
\end{figure}

A model H${\alpha}$ emission line profile is constructed by employing a Monte Carlo simulation of 
a spherically symmetric distribution\footnotemark \footnotetext{As noted in D11 details of the
resulting line profile shape are sensitive to the polar and azimuthal angles ${\theta}$
and ${\phi}$. Restricting the range of ${\theta}$ such that -0.97${\pi}$/2~${\le}~{\theta}~{\le}~0.97{\pi}$/2 simulates the cavity expected to be occupied by the radio jets and causes the model profile to more closely mimic the observed profile at zero velocity.}
of ${\sim}$10$^4$ points of light, the radial distribution of which is described by the H${\alpha}$ emission line emissivity. The central mass determines the relationship between velocity and radius for each point of light, and the emissivity determines the number of points at each radius. Thus, knowing both the H${\alpha}$ emission line emissivity and the central mass allows one to construct 
a model H${\alpha}$ emission line profile given a kinematic description for the BLR gas.
Regrettably, the kinematic state of the BLR gas is unknown.
However, for the purposes of computing model line profiles, every point of light was assumed to be moving at the escape velocity, which is equivalent in magnitude to the free-fall velocity, according to the familiar equation ${v(r)}$ = ${\sqrt{ 2 G M(r)/r}}$, where ${v}$ is velocity, ${G}$ is the gravitational constant, ${M(r)}$ is the mass interior to ${r}$, where ${r}$ is the radial distance of each point from the central supermassive black hole (BH). The radial dependence of the central mass arises because ${M(r)}$ includes both the BH mass ${M_{\bullet}}$, and the surrounding stars ${M_{\star}(r)}$, which will be important if the BLR gas is spatially extended 
as appears to be the case for NGC 3998.
Such spherically symmetric free-fall models are able to reproduce the distinctly triangular shape observed for the broad H Balmer lines in NGC 3998, whereas a thin accretion disc would produce two emission peaks unless the disc is contrived to be nearly face-on (D11).
Values adopted for ${M_{\star}(r)}$ are listed in Table 1 and represent the same stellar mass model employed
by \citet[][M. Sarzi 2016, private communication]{Walsh2012a} and a distance to NGC 3998 of 14 Mpc \citep{Tonry2001}. Given the likelihood that the BLR gas is spatially extended, no contribution to the model emission line profile was allowed for points of light that lie beyond the angular extent of the STIS slit, corresponding to 52${\arcsec} {\times} 0.1{\arcsec}$. The goodness of fit between the model H${\alpha}$ line 
profile and the observed one is measured using the reduced chi-squared statistic
\begin{equation}
{\chi}_{red}^2 = {\sum_j} (O_j - M_j)^2/(\nu \delta^2)
\end{equation}
where ${O_j}$ represents the observed normalized line profile intensities,  ${M_j}$, the model normalized line profile intensities, ${\nu}$ = 470 degrees of freedom
and ${\delta}$ the uncertainty in the observed normalized line profile intensities, taken to be 4\%. The summation was performed over the velocity span of the broad H${\alpha}$ line depicted in Figure 2.

\begin{table}
\centering
\caption{Stellar Mass Model of NGC 3998}
\begin{tabular}{cccc}
\hline
Radius  & Radius & Velocity & M$_{\star}$    \\
${\arcsec}$ & pc &  km/s & M${_{\odot}}$ \\
(1) & (2) &  (3)  & (4)  \\
\hline
1.0 ${\times}$ 10$^{-1}$ & 7 & 356.4 & 2.0 ${\times}$ 10$^{8}$  \\
8.7 ${\times}$ 10$^{-2}$ & 6 & 358.4 & 1.7 ${\times}$ 10$^{8}$ \\
7.3 ${\times}$ 10$^{-2}$ & 5 & 365.1 & 1.5 ${\times}$ 10$^{8}$ \\
5.8 ${\times}$ 10$^{-2}$ & 4 & 382.4 & 1.3 ${\times}$ 10$^{8}$ \\
4.4 ${\times}$ 10$^{-2}$ & 3 & 406.3 & 1.3 ${\times}$ 10$^{8}$ \\
2.9 ${\times}$ 10$^{-2}$ & 2 & 393.3 & 6.9 ${\times}$ 10$^{7}$ \\
1.4 ${\times}$ 10$^{-2}$ & 1 & 261.9 & 1.5 ${\times}$ 10$^{7}$ \\
\hline
\end{tabular}
\end{table}

\subsection{Photoionization Modelling Results}
The earliest photoionization models of Seyferts and Liners \citep{Ferland1983,Halpern1983} implicated low metallicity gas. Subsequently, \cite{Binette1985} noted the connection between the shape of the ionizing continuum and the metal abundance inferred from the [O\,{\sc iii}]${\lambda}$5008/H${\beta}$ ratio. As explained in more detail in the following, for the ADAF continuum of \cite{Nemmen2014}, a low metallicity is required to reproduce the [O\,{\sc iii}]${\lambda}$5008/H${\beta}$ ratio observed with STIS (D11).

A grid of photoionization models spanning a range of metallicity relative to solar $Z/Z_\odot$, where 10$^{-3}$ ${\leq}$  $Z/Z_\odot$ $ {\leq}$ 10$^{-1}$, power law index, ${n}$, where 
0.5 ${\leq}$  ${n}$ ${\leq}$ 1.25, and gas density ${\rho}_o$, where 4 ${\leq}$ log${_{10}}$${\rho_o}$(cm${^{-3}}$) ${\leq}$ 5, was constructed
in order to discover the intersection of model predictions with the observed reddening-insensitive
[O\,{\sc iii}]${\lambda}$5008/H${\beta}$ and [O\,{\sc iii}](${\lambda}$5008 + ${\lambda}$4960)/${\lambda}$4364 ratios.\footnotemark \footnotetext{All reported wavelengths are vacuum.} Additionally, each model produced a unique Balmer line emissivity which was used to generate
a model H${\alpha}$ emission line profile shape
for comparison with a normalized version of the observed one, as described previously in Section 2.1. 

\begin{figure}
	\includegraphics[width=\columnwidth]{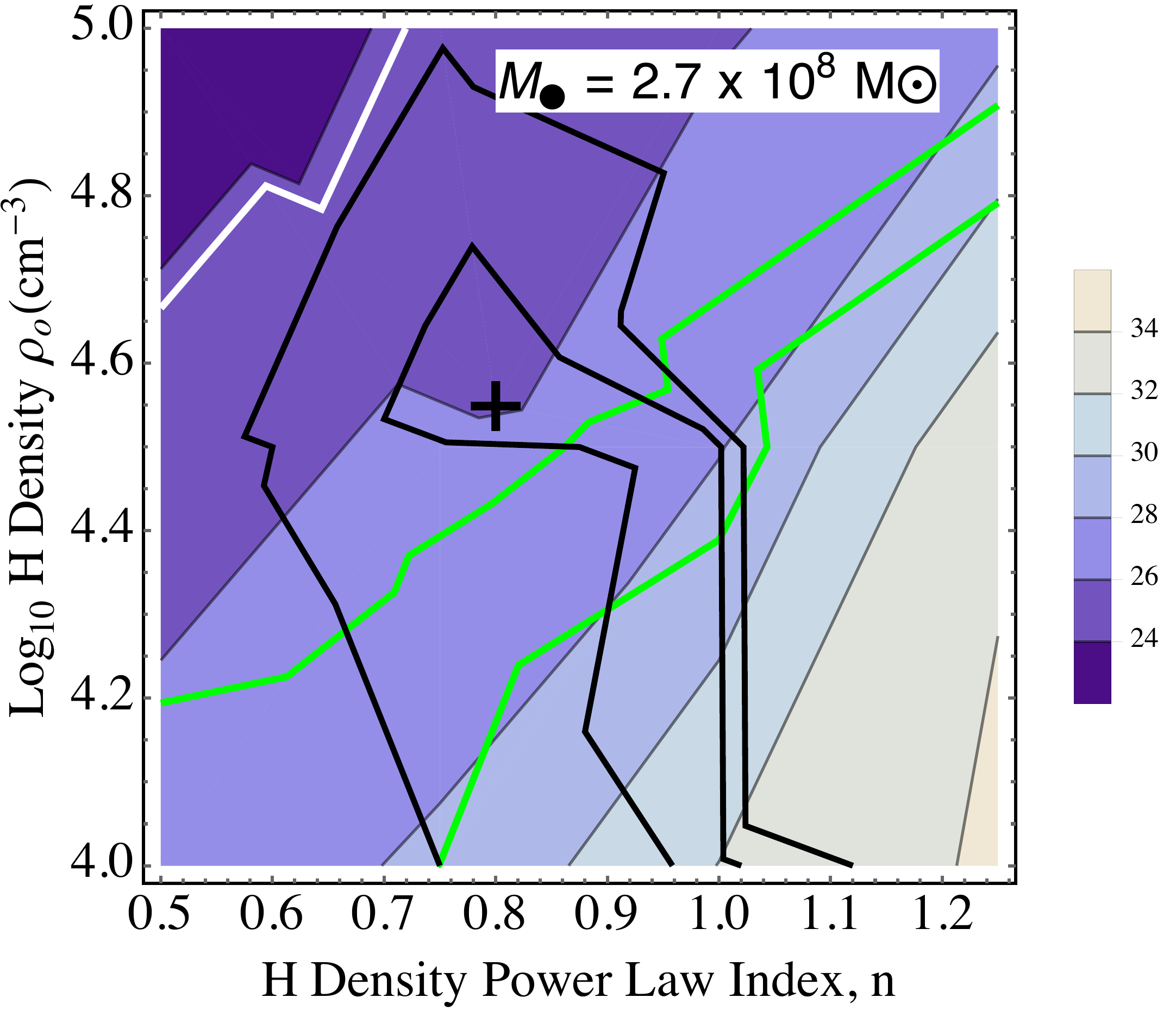}
    \caption{Photoionization model results as a function of gas density ${\rho}_o$ and index $n$ for a metallicity $Z/Z_\odot$ = 0.01.
The ordinate identifies the H density ${\rho}_o$ at a reference radius $r_o$ = 2.38 ${\times}$ 10$^{17}$ cm. The abscissa identifies the index $n$ of the power law used to describe the radial distribution of neutral H gas (see Section 2 for details).
The intersection of 
the model [O\,{\sc iii}]${\lambda}$5008/H${\beta}$ ratio with the observed value is depicted by the 
green lines that bound the ${\pm}$ 1${\sigma}$ uncertainties. Lines of constant ionization parameter run parallel to the abscissa.
The intersection of 
the model [O\,{\sc iii}](${\lambda}$5008 + ${\lambda}$4960)/${\lambda}$4364 ratio with the observed value, a lower limit, is depicted by the white line. The legend refers to the background coloured contours depicting the [O\,{\sc iii}](${\lambda}$5008 + ${\lambda}$4960)/${\lambda}$4364 ratio, a measure of electron temperature.  Black lines identify ${\chi}_{red}^2$ contours representing the goodness of fit between the model H${\alpha}$ emission line profile and the observed one (see Figure 2 and Section 2.1)  for the BH mass 2.7 ${\times}$ 10${^{8}}$ M${_{\sun}}$ measured by \citet{Francesco2006}. Two contours are plotted representing ${\chi}_{red}^2$ = 2 and ${\chi}_{red}^2$ = 3. 
The black cross represents ${\chi}_{red}^2$ = 1.2 and the corresponding model H${\alpha}$ emission line profile is illustrated in Figure 2. }
    \label{fig:example_figure}
\end{figure}

Model results for the grid of photoionization models described above are displayed in Figure 3 as a function of gas density ${\rho}_o$ and index $n$ 
for a metallicity $Z/Z_\odot$ = 0.01. Such a low metallicity is required for the model to reproduce the 
observed [O\,{\sc iii}]${\lambda}$5008/H${\beta}$ ratio = 0.65 ${\pm}$ 0.03 (D11), depicted by the green lines in Figure 3. Incidentally, that ratio
qualifies the AGN in NGC 3998 as a Liner 1.2 according to the type scheme proposed for Seyferts by \cite{Whittle1992}. Background shaded contours in Figure 3 refer to the [O\,{\sc iii}](${\lambda}$5008 + ${\lambda}$4960)/${\lambda}$4364 ratio, a measure of electron temperature. 

An interesting and important result to have emerged from these photoionization models is that the [O\,{\sc iii}]${\lambda}$5008/H${\beta}$ ratio depends on metallicity $Z/Z_\odot$ when the ionization parameter ${\xi}$ or U is held constant. The result is illustrated in Figure 4 where [O\,{\sc iii}]${\lambda}$5008/H${\beta}$ = 60 $Z/Z_\odot$ for a representative row of points in Figure 3 for which log$_{10} {\xi}$ = 2.7. Other model parameters such as the [O\,{\sc iii}](${\lambda}$5008 + ${\lambda}$4960)/${\lambda}$4364 ratio which measures the electron temperature, the H${\alpha}/$H${\beta}$ ratio which is a surrogate for electron density \citep{Devereux2016}, the H${\beta}$
luminosity, L$_{H{\beta}}$ a measure of the number of ionizing photons, and ${{\chi}_{red}^2}$ a measure of the goodness of fit between the model and observed H${\alpha}$ emission line profile, hardly depend on metallicity, if at all. With hindsight one can see that the [O\,{\sc iii}]${\lambda}$5008/H${\beta}$ ratio tracks metallicity at fixed $U$ in the photoionization models of \citet[][see their Figure 2]{Ferland1983}. However, the present work shows that the dependence of [O\,{\sc iii}]${\lambda}$5008/H${\beta}$ on metallicity persists for a range of ionization parameters,  2.2 ${\leq}$  log$_{10} {\xi}$ ${\leq}$ 3.2, that are many orders of magnitude higher than typically employed in AGN photoionization models.
In the context of the present analysis, the principal utility of the correlation is that the [O\,{\sc iii}]${\lambda}$5008/H${\beta}$ ratio very precisely constrains the metallicity of the photoionized gas in NGC 3998 to be 0.01$Z/Z_\odot$, as illustrated in Figures 3 and 4. Although not shown, if Figure 3 were reproduced for metallicities bracketing 0.01$Z/Z_\odot$, for example 10$^{-3}Z/Z_\odot$ and 10$^{-1}Z/Z_\odot$, the figures would look virtually identical to Figure 3 except that the green lines would not be present because the [O\,{\sc iii}]${\lambda}$5008/H${\beta}$ ratio coincides with the observed one only for 0.01$Z/Z_\odot$.

\begin{figure}
	\includegraphics[width=\columnwidth]{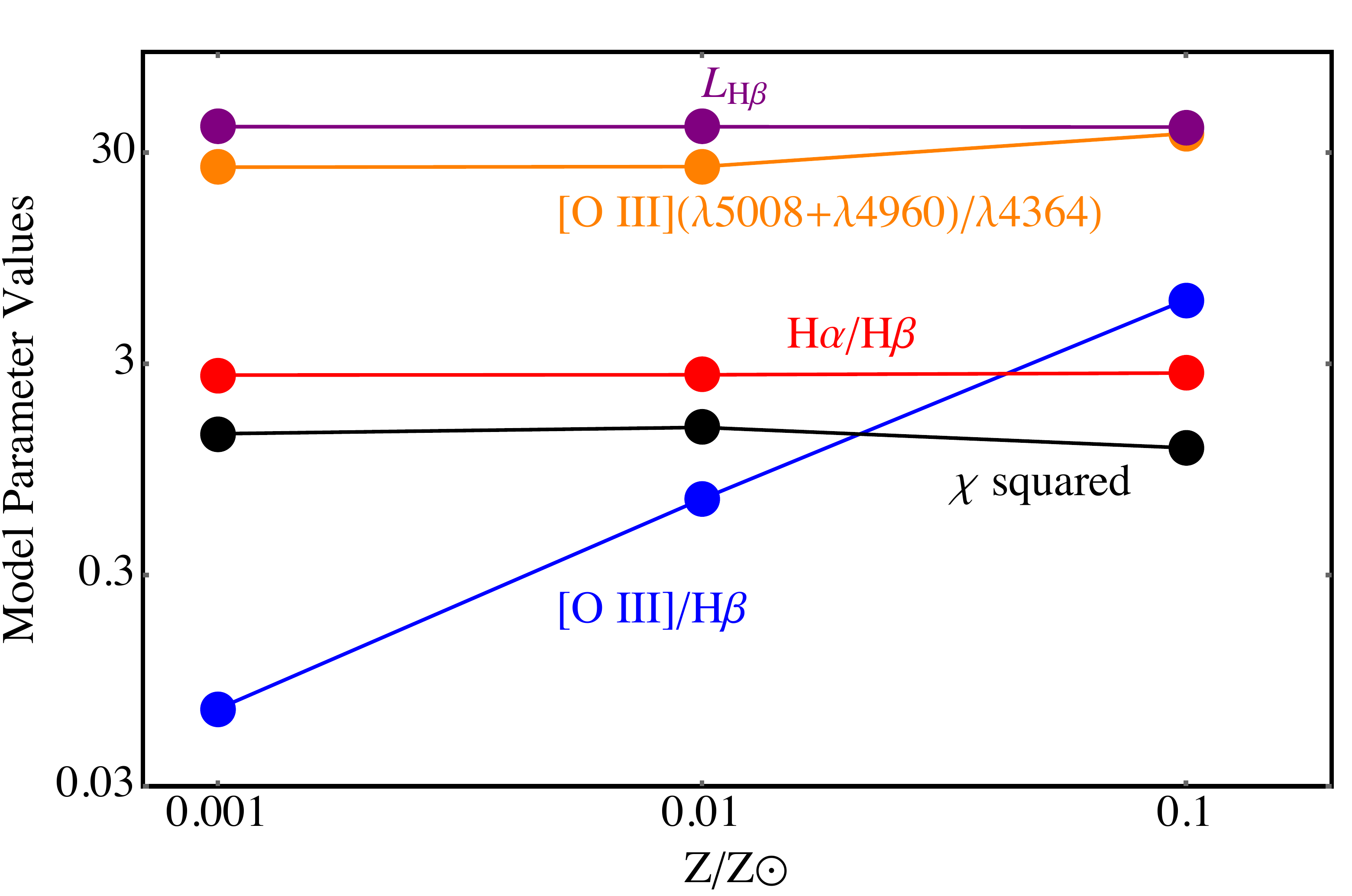}
    \caption{The dependence of various model parameter values on metallicity $Z/Z_\odot$ for fixed ionization parameter log$_{10} {\xi}$ = 2.7. The purple line represents log$_{10}$ of the H${\beta}$
luminosity, L$_{H{\beta}}$ in erg s$^{-1}$, the orange line represents the [O\,{\sc iii}](${\lambda}$5008 + ${\lambda}$4960)/${\lambda}$4364 ratio, a measure of electron temperature, the red line H${\alpha}/$H${\beta}$ a surrogate of electron density, the black line ${{\chi}_{red}^2}$, a measure of the goodness of fit between the model and observed H${\alpha}$ emission line profile, and the blue line represents the [O\,{\sc iii}]${\lambda}$5008/H${\beta}$ ratio which depends linearly on metallicity.
}
    \label{fig:example_figure}
\end{figure}

A feature of the models is that a wide range of ${\rho}_o$,  ${n}$ combinations produce model H${\alpha}$ emission line profiles that mimic the observed one (Figure 2) portrayed by the elongated black contours extending towards low gas densities in Figure 3. Careful examination of these models reveals a degeneracy that includes a wide range of
possible sizes for the H$^+$ region. However, the predicted ionization structure is such that the outer radius of the H$^+$ region coincides
with the location of the 
O$^{++}$ region as illustrated in the upper panel of Figure 5. In this sense the ionization structure is similar to that described by \citet{Osterbrock1989} for a planetary nebula, suggesting a way to break the size degeneracy amongst these various models using the observed [O\,{\sc iii}]${\lambda}$5008
emission line. A useful observational constraint is that 100\% of the average [O\,{\sc iii}]${\lambda}$5008
flux measured with ground-based telescopes \citep{Heckman1980,Blackman1983a,Ho1997} originates from a region with an angular size comparable to the 0.2${\arcsec}$ STIS slit width employed for the G430L observations. Consequently, the size of the O$^{++}$ region, and by association the H$^+$ region,
must be ${\leq}$ 0.2${\arcsec}$ a constraint that is satisfied, coincidentally, by models that lie above the upper of the two green lines in Figure 3. Thus, a plausible solution, identified by the black cross in Figure 3, that satisfies all three reddening independent observables, corresponds to a gas density ${\rho}_o$ = 3.55 ${\times}$ 10$^{4}$ cm$^{-3}$ at $r_o$ = 2.38 ${\times}$ 10$^{17}$ cm and a power law index $n$ = 0.8 for the BH mass, 2.7 ${\times}$ 10${^{8}}$ M${_{\sun}}$, measured using gas kinematics by \citet{Francesco2006}. No compelling solutions were found for the BH mass, 8.1 ${\times}$ 10${^{8}}$ M${_{\sun}}$, measured using stellar kinematics \citep{Walsh2012a} principally because that BH mass produced model H${\alpha}$ emission line profiles that consistently did not match the observed one as illustrated by the much larger values of ${\chi}_{red}^2$ in Figure 6, compared to Figure 3.

\begin{figure}
	\includegraphics[width=\columnwidth]{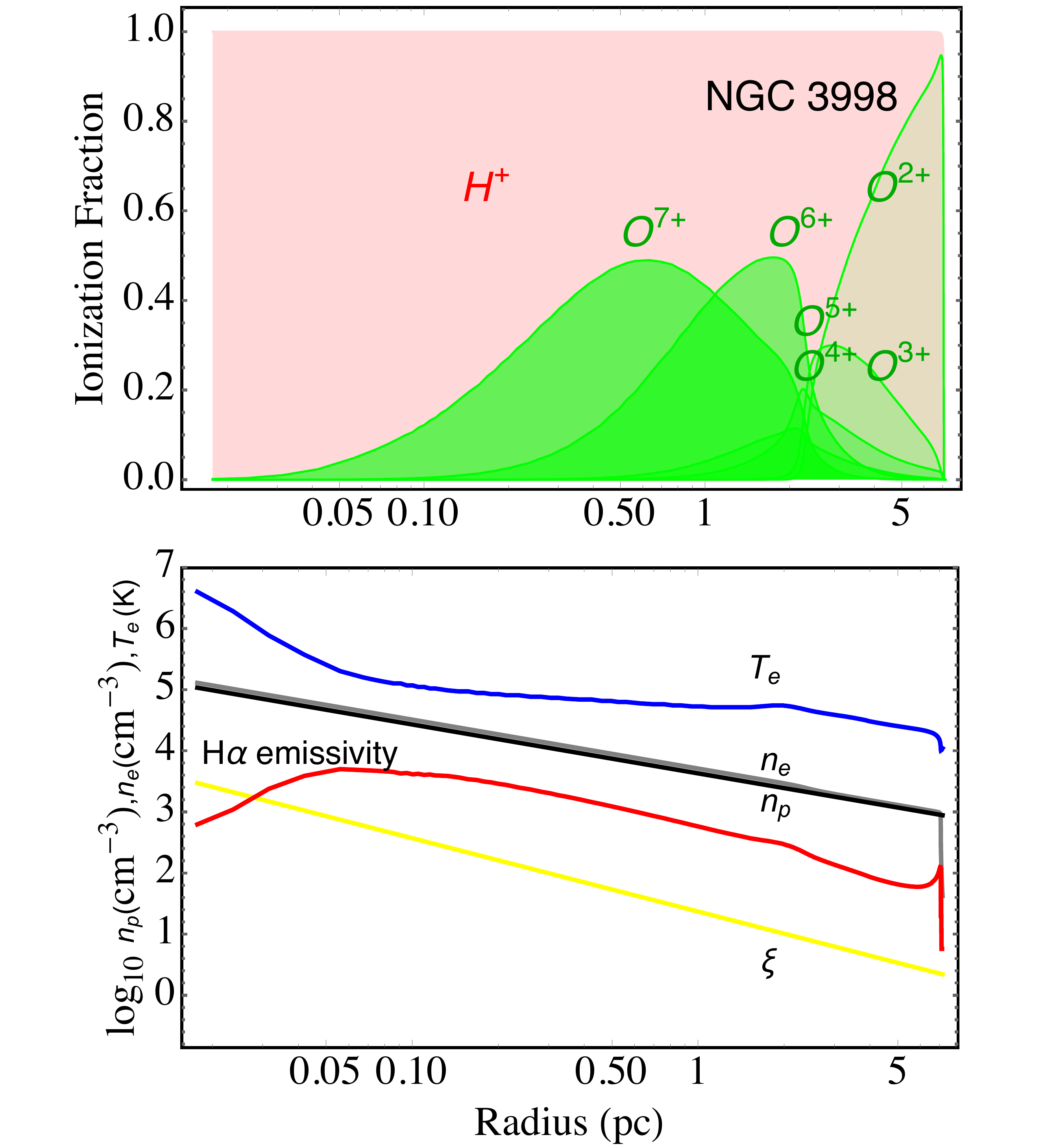}
    \caption{Radial distributions for the model parameters ${\rho}_o$ = 3.55 ${\times}$ 10$^{4}$ cm$^{-3}$ at $r_o$ = 2.38 ${\times}$ 10$^{17}$ cm, $n$ = 0.8, $Z/Z_\odot$ = 0.01. The upper panel shows the ionization fractions of H${^+}$ (pink shading) and all possible ionization stages of O (green shading). The lower panel illustrates the electron temperature, T$_e$(K)  (blue line), electron density, n$_e$ (cm$^{-3}$) (grey line), proton density, n$_p$ (cm$^{-3}$) (black line), the H${\alpha}$ emissivity in arbitrary units (red line) and the ionization parameter ${\xi}$ (yellow line). The ordinate refers to a variety of units whereas the abscissa indicates distance from the BH in pc.}
    \label{fig:example_figure}
\end{figure}

\begin{figure}
	\includegraphics[width=\columnwidth]{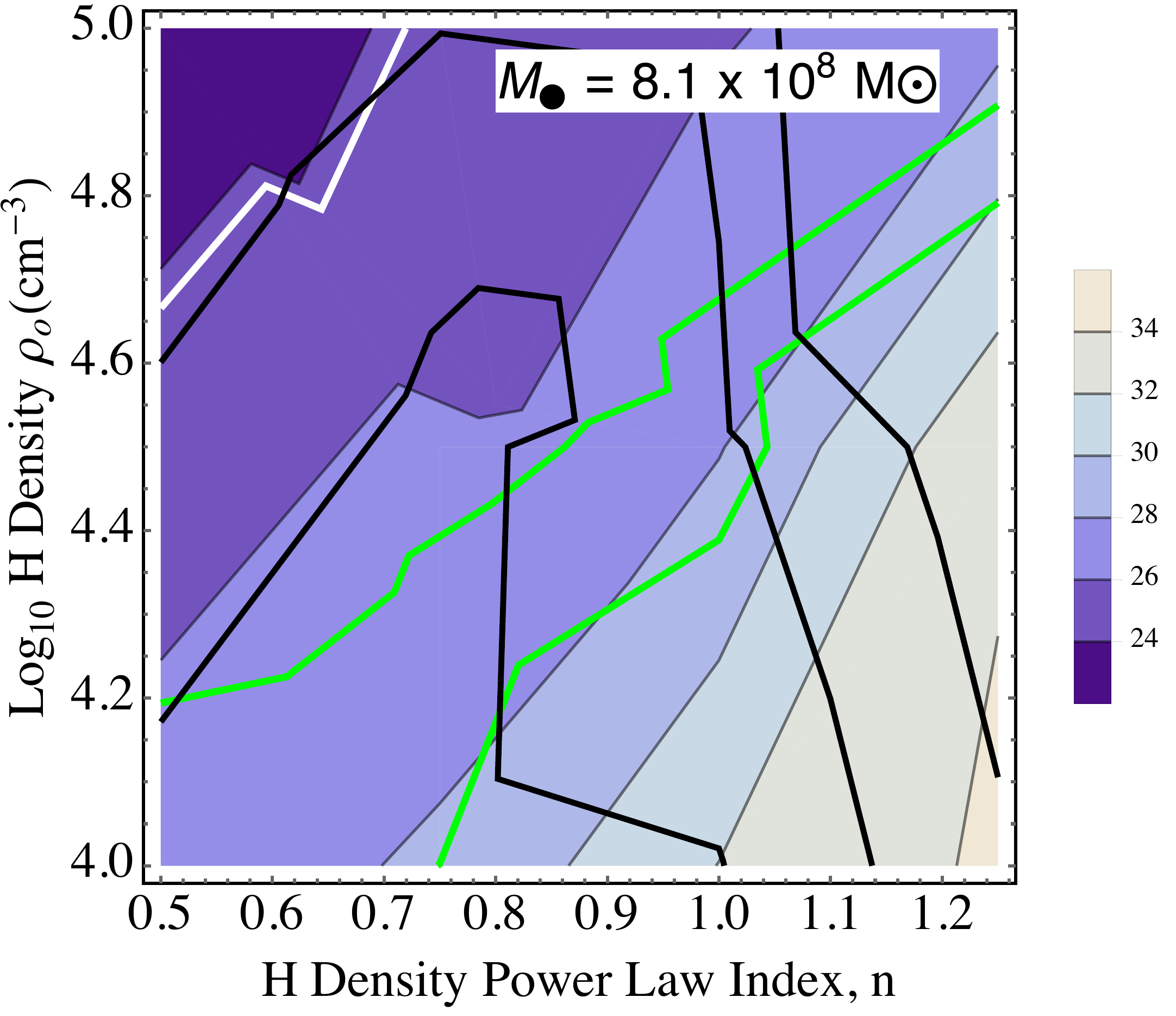}
    \caption{Photoionization model results as a function of gas density ${\rho}_o$ and index $n$ for a metallicity $Z/Z_\odot$ = 0.01. Black lines identify ${\chi}_{red}^2$ contours representing the goodness of fit between the model H${\alpha}$ emission line profile and the observed one (see Figure 2 and Section 2.1) for the BH mass, 8.1 ${\times}$ 10${^{8}}$ M${_{\sun}}$, measured by \citet{Walsh2012a}. Three contours are plotted representing ${\chi}_{red}^2$ = 10, 20 and 30, respectively. Otherwise, the figure and associated legend 
are identical to Figure 3.}
    \label{fig:example_figure}
\end{figure}

The physical dimensions of the BLR in NGC 3998, synonymous now with the H$^+$ region, are determined by the H${\alpha}$ emissivity, 
the radial dependence of which is illustrated in the lower panel of Figure 5. As the UV--X-ray source is approached, the electron
temperature increases rapidly at a radius ${\sim}$0.05pc, where the H${\alpha}$ emissivity concurrently declines because the H gas is now an X-ray emitting plasma, for which the primary source of opacity is electron scattering. This turnover in the H${\alpha}$ emissivity at small radius provides a natural explanation for the 
finite width observed for the Balmer lines. The outer radius of the BLR is large, ${\sim}$7 pc, corresponding to a sharp decrease in the H${\alpha}$ emissivity that coincides with the outer boundary of the H${^+}$ region at a H equivalent column density of 6.5 ${\times}$ 10$^{22}$ atoms cm$^{-2}$. That the H${\alpha}$ emissivity provides a natural explanation for the
overall dimensions of the BLR makes the H${\alpha}$ emission line profile model presented in Section 2.1 all the more  compelling, especially when compared to other BLR models in the published literature that 
require the inner and outer radii to be parameterized.

The upper limit of 28,800 K for the electron temperature constrained using the
observed reddening-insensitive [O\,{\sc iii}]${\lambda}$${\lambda}$${\lambda}$5008.24,4960.30,4364.44 lines (D11)
is also consistent with the temperature predicted for the O$^{++}$ zone (see Figure 5).
In summary, the photoionization model presented for NGC 3998 in Figure 5 provides a natural and straightforward physical explanation for the dimensions of the BLR,
and hence the shape of the H${\alpha}$ emission line profile, as well as the electron temperature in the O$^{++}$ zone. 

\subsection{Dust Extinction to the BLR and the Central UV Continuum Source.}

Galactic extinction to NGC 3998 is negligible, but there is undoubtedly some dust extinction to the Liner nucleus of NGC 3998 caused by the host galaxy \citep{Eracleous2010}. However, dust extinction internal to the BLR gas must be small at visible wavelengths because the Balmer emission line profile shapes are so similar (D11).

The foreground dust extinction ${A_\lambda}$ may be computed for every observed wavelength ${\lambda}$ by comparing the observed flux ${f_{\lambda}}$ with the intrinsic flux ${F_{\lambda}}$,

\begin{equation}
A_{\lambda} = 2.5 log\frac{F_{\lambda}}{f_{\lambda}}~~~~\rm{mag}
\end{equation}
Subsequently, expressing the total extinction ${A_\lambda}$,
as the function of two fixed parameters (${A_{\infty},b}$) and one variable, ${\lambda }$, \citep{Cardelli1989,Whitford1958}, so that
\begin{equation}
A_{\lambda} = A_\infty + \frac{b}{\lambda} ~~~~\rm{mag}
\end{equation}
allows one to identify the constant ${A_{\infty}}$ with the extinction at infinite wavelength, which is expected to be zero. Examples of extinction curves derived using this methodology are presented in Figure 7.

\begin{figure}
	\includegraphics[width=\columnwidth]{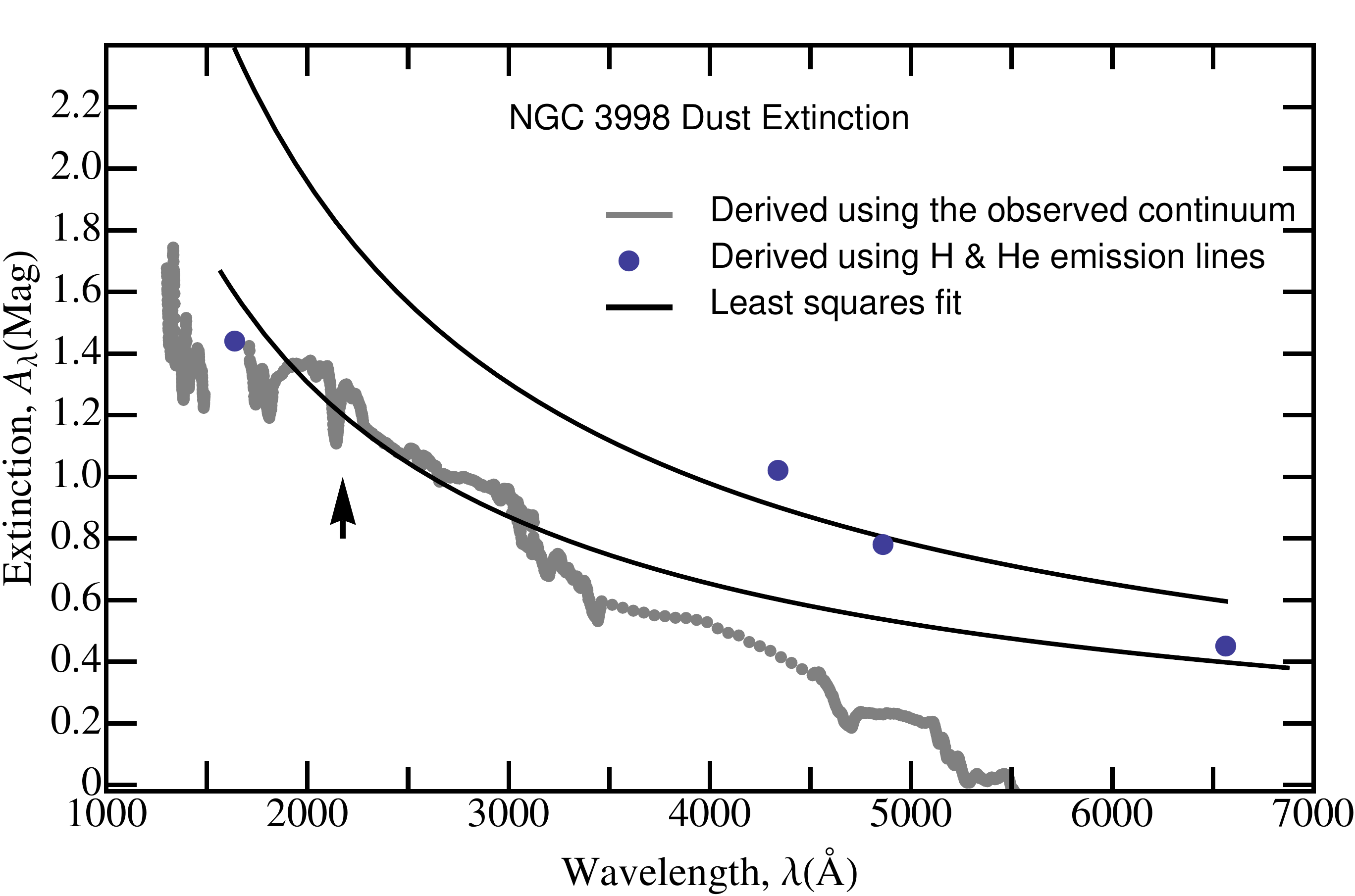}
    \caption{Rest frame extinction to the emission line-free UV--visible continuum in NGC 3998 is defined by comparing contemporaneous STIS observations made with the G140L, G230L, G430L and G750M gratings to the intrinsic ADAF continuum of \citet{Nemmen2014} (light grey line). The grey dots centred at 4000 {\AA} represent an interpolation of the observed continuum over the gap where several emission lines have been removed.
Extinction to the H${^+}$ region is defined by comparing contemporaneous STIS observations of the H${\alpha}$, H${\beta}$, H${\gamma}$ and ${\lambda}$1640 ${\ion{He}{ii}}$ emission lines with the intrinsic
emission line fluxes predicted by XSTAR (see Table 3). The ordinate indicates total dust extinction in units of magnitudes.  The abscissa indicates wavelength in the rest frame of NGC 3998 expressed in units of {\rm \AA}, and the black arrow identifies the wavelength expected for the ${\lambda}$2175${\rm \AA}$ feature.} 

    \label{fig:example_figure}
\end{figure}

The continuum extinction is derived by adopting the ADAF continuum of \cite{Nemmen2014} as the intrinsic continuum and the
contemporaneous STIS observations obtained with the G140L, G230L, G430L and G750M gratings, following removal of the emission lines, as
the observed continuum. 

The emission line extinction is derived by adopting model predictions for the intrinsic emission line luminosities produced by the ${\rho}_o$ = 3.55 ${\times}$ 10$^{4}$ cm$^{-3}$ at $r_o$ = 2.38 ${\times}$ 10$^{17}$ cm, $n$ = 0.8, $Z/Z_\odot$ = 0.01 model, which are then compared to the observed H${\alpha}$, H${\beta}$, and H${\gamma}$ 
emission line fluxes taken from D11, plus the ${\lambda}$1640 ${\ion{He}{ii}}$ emission line luminosity which is presented for the first time in Table 3. 

\begin{table}
\centering
\caption{Dust Extinction to Bright H \& He Emission Lines}
\begin{tabular}{cccc}
\hline
Line  & Observed & XSTAR & Extinction   \\
 & 10$^{39}$ erg/s & 10$^{39}$ erg/s & mag \\
(1) & (2) &  (3)  & (4)  \\
\hline
H${\alpha}$ & 10.6 & 16.0 & 0.45 \\
H${\beta}$ &  2.9 & 6.0 & 0.78 \\
H${\gamma}$ & 1.0 & 2.6 & 1.03 \\
${\lambda}$1640 ${\ion{He}{ii}}$ & 1.2 & 4.4 & 1.41 \\
\hline
\end{tabular}
\end{table}

Given that the measurement uncertainty is about the same as the point size used in Figure 7, the extinction to the UV--visible continuum is 
significantly different from that derived using the three brightest Balmer emission lines. 
An unweighted least squares fit to the H emission lines yields the following extinction law:
\begin{equation}
A_{\lambda} = \frac{(3912 \pm 390)}{\lambda(\textrm\AA)}~~~~\rm{mag}
\end{equation}
represented by the upper of the two black curves plotted in Figure 7. According to this extinction law, ${E(B-V)}$ = 0.17 ${\pm}$ 0.01, slightly higher than the value
estimated by \cite{Eracleous2010} based on the H column required to explain the X--ray absorption. 

Interpretation of the UV--visible continuum extinction is more complicated because significant structure exists in the derived extinction. Experimentation revealed that a least squares fit to just the G230L continuum yields the most statistically significant result:
\begin{equation}
A_{\lambda} = \frac{(2612 \pm 9)}{\lambda(\textrm\AA)}~~~~\rm{mag}
\end{equation}
represented by the lower of the two black curves plotted in Figure 7 and for which ${A_{\rm v}/E(B-V)}$= 4.3 ${\pm}$ 0.4, in the realm of
values observed for the Galaxy \citep{Cardelli1989}. 

The visible extinction derived using the G430L and G750M continuum observations falls below an extrapolation of the least squares fit to the G230L extinction expressed in Equation 8. This is most likely due to an increasing contribution to the observed visible continuum from starlight at longer wavelengths as noted previously by \cite{Nemmen2014}. Another interesting observation, illustrated in Figure 7, is that the extinction derived using both the ${\lambda}$1640 ${\ion{He}{ii}}$ emission line and the observed G140L 
continuum flattens in the far UV, ${\lambda}~{\leq}~2000{\rm \AA}$. Such greying of the dust opacity has been noted for other AGN as has the absence of
the ${\lambda}$2175${\rm \AA}$ feature \cite[][and references therein]{Gaskell2004,Gaskell2017a}.

Finally, XSTAR predicts that aside from ${\lambda}$1640 ${\ion{He}{ii}}$, the brightest UV emission lines 
produced inside the H${^+}$ region are Ly${\alpha}$, ${\lambda}$1548 ${\ion{C}{iv}}$,
and ${\lambda}$1906 ${\ion{C}{iii}}$, consistent with the STIS observations presented in D11, although 
a more detailed analysis is beyond the scope of the present paper, largely because 
of the absorption associated with Ly${\alpha}$ and ${\lambda}$1548 ${\ion{C}{iv}}$.

\section{Discussion}

H is the most abundant element in the BLR of NGC 3998 and must be the cornerstone of any viable photoionization model. The H${\alpha}$ emission line, in particular, is sufficiently well resolved in velocity space to pursue sensitive kinematic modelling of the line profile shape using emissivities computed by XSTAR. Photoionization modelling with XSTAR demonstrates that the H Balmer, [O\,{\sc iii}] and possibly several of the broad UV emission lines seen in the spectrum of the LINER 1.2 nucleus of NGC 3998 can be explained in terms of a large volume of low density, low metallicity gas that is photoionized by the central UV--X-ray source. The evidence for low metallicity is supported by broad Balmer emission line profile shapes that are observed to be symmetrical and similar to each other (D11) indicating that the H${^+}$ region must be devoid of ${\micron}$-sized dust grains. These insights result from a novel approach
whereby three reddening-insensitive observables, namely the [O\,{\sc iii}]${\lambda}$5008/H${\beta}$ and [O\,{\sc iii}](${\lambda}$5008 + ${\lambda}$4960)/${\lambda}$4364 ratios, plus the H${\alpha}$ emission line profile shape,
are used to constrain XSTAR photoionization models which can then be used to infer dust extinction.
Usually, observers do the reverse, they attempt to de-redden emission line ratios before using them to constrain photoionization models, but the pitfall with that approach is identifying a suitable reddening indicator at each observed wavelength.  The H${\alpha}/$H${\beta}$ ratio is often used e.g. \citet[and references therein]{Heard2016, Gaskell2017a},
but that is a troublesome diagnostic because the ratio is also a proxy for electron density in some AGN \citep{Devereux2016}. Although, having said that, the H${\alpha}/$H${\beta}$ ratio does appear to be a measure
of reddening in NGC 3998 as Figure 7 illustrates.

\subsection{Does NGC 3998 have a BLR?}

NGC 3998 certainly has broad emission lines, but does it have a BLR?  Answering this question requires a definition. \cite{Laor2003} proposes that the BLR is the region interior to the dust sublimation radius. Although a perfectly reasonable definition, in the case of NGC 3998 it is inadequate for two reasons. First, energy balance considerations \citep{Barvainis1987} indicate that the luminosity of the central UV--X-ray source is insufficient to sublimate dust anywhere inside the H${^+}$ region responsible for the broad Balmer emission lines. Second, the striking similarity between the H${\alpha}$, H${\beta}$ and H${\gamma}$ emission lines reported by D11 suggests that the broad Balmer lines arise from a photoionized region that is devoid of dust, presumably because of an intrinsically low metallicity rather than dust sublimation.
A more rigorous definition, proposed here, is that the BLR is synonymous with the H${^+}$ region from where the broad Balmer emission lines emerge. That definition naturally entails an explanation for the physical dimensions of the BLR, at least in NGC 3998, in terms of the Balmer emissivity as explained previously in Section 2.2. 

\begin{figure}
	\includegraphics[width=\columnwidth]{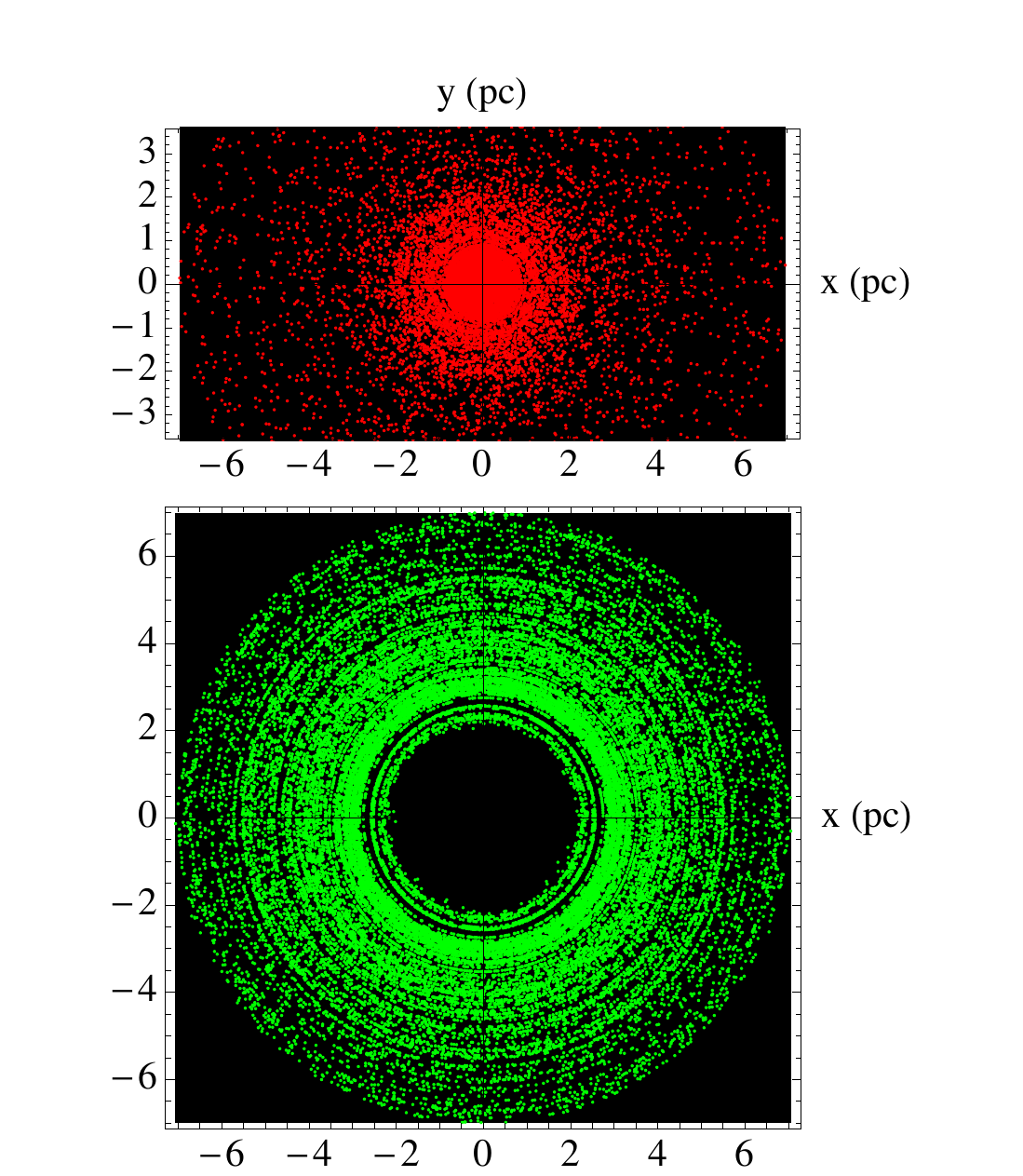}
    \caption{Visualization of the BLR in NGC 3998. Top panel: the ordinate depicts the width, in pc, of the STIS G750M slit used to obtain the H${\alpha}$ observations. Red dots represent the same distribution of points that produced the model H${\alpha}$ emission line illustrated in Figure 2. Lower panel: the ordinate depicts the width, in pc, of the STIS G430L slit used to obtain the [O\,{\sc iii}] observations. The green dots represent the [O\,{\sc iii}]${\lambda}$5008 emissivity. The abscissa illustrates the radial extent of the BLR in both plots in units of pc.
}
    \label{fig:example_figure}
\end{figure}

NGC 3998 also has broad [O\,{\sc iii}]${\lambda\lambda}$5008,4960 emission lines. \cite{Laor2003} anticipated such an observation if the BLR size-luminosity correlation extends to low luminosity AGN like NGC 3998. Then, as the BLR size decreases, the velocity width of the Balmer lines increases. Because the [O\,{\sc iii}] emission lines are produced in a region that encircles the BLR
they may also broaden by association. The XSTAR results do indeed produce an ionization structure for the AGN in NGC 3998 that is similar to that envisaged by \cite{Laor2003} as Figure 8 illustrates, but 
those same results also indicate that the BLR is several orders of magnitude larger than one would expect based on the BLR size-luminosity relationship.
The most
straightforward interpretation of this result is that the BLR size-luminosity relationship established for reverberating AGN does not apply to non-reverberating AGN like NGC 3998  \citep{Devereux2015}. Part of the size discrepancy results from the observation that the reverberation radius measures only the inner radius of a much larger volume of ionized H gas \citep[and references therein]{Devereux2016}. However, even the inner radius of the Balmer emitting region in NGC 3998 (${\sim}$0.05 pc, see Section 2.2 and Figure 5) is two orders of magnitude larger than expected. Furthermore,
if stars do contribute to the nuclear luminosity in the visible, as the results presented in Section 2.3 indicate, the corrected luminosity will be smaller which will only exacerbate the size discrepancy reported for NGC 3998 by \citet{Devereux2015}. Based on the results presented here for NGC 3998, compared to a similar study
of the low luminosity Seyfert 1 nucleus in NGC 3516 \citep{Devereux2016}, one can anticipate the density of gas in the BLR to be an important parameter distinguishing non-reverberating AGN from their reverberating counterparts.

\subsection{Comparison of  H${^+}$ and LOC Models}

The XSTAR code is designed to compute the emergent spectrum for a spherical H${^+}$ region photoionized by a centrally
located UV--X-ray source. The resulting ionization structure is such that little or no emission is predicted for low ionization forbidden lines in contradiction to the observed STIS spectrum which clearly shows [S\,{\sc ii}], [O\,{\sc ii}] and [O\,{\sc i}] with a full width half maximum velocity that matches the Balmer lines (D11). Regrettably, these forbidden lines are too faint to model, and so the location of the low ionization gas can not be constrained currently using the methodology described in this paper. However, it is quite clear that the low ionization gas must be located outside the H${^+}$ region. Thus, according to this model there are possibly two separate components,
a highly ionized ball of mostly H surrounded by gas of lower ionization and possibly higher metallicity. Interestingly, the notion of two components was first introduced
by \cite{Heckman1980} to explain a curious observation that [O\,{\sc iii}] emission is seemingly independent of both [O\,{\sc ii}] and [O\,{\sc i}] in Liners and Seyferts.

The model presented here for NGC 3998 predicts the shape of the broad H${\alpha}$ emission line, as well as the relative intensities and luminosities for the H Balmer, [O\,{\sc iii}], and potentially several of the broad UV emission lines at the expense of just three free parameters, ${\rho}_o$, $r_o$ and $Z/Z_\odot$. This simplicity is to be contrasted with local optimally emitting cloud (LOC) photoionization models of Liners and Seyfert 2s which invoke an inhomogeneous interstellar medium characterised by an ensemble of clouds of solar or super-solar metallicity, low ionization parameter, each cloud spanning a wide range of density that is independent of galactocentric radius \citep{Komossa1997, Ferguson1997, Richardson2014}. Needless to say, this approach to photoionization modelling is completely different because LOC models aim to explain narrow emission lines observed in low angular resolution spectra which display a wide range of ionization.

However, LOC models do have a long-standing problem in that they consistently over-predict the [O\,{\sc iii}](${\lambda}$5008 + ${\lambda}$4960)/${\lambda}$4364 ratio in low ionization AGN \citep{Komossa1997, Ferguson1997, Richardson2014} whereas the best fitting model for the H${^+}$ region in the Liner 1.2 nucleus of NGC 3998 predicts an [O\,{\sc iii}](${\lambda}$5008 + ${\lambda}$4960)/${\lambda}$4364 ratio that is just 8\% higher than the observed lower limit\footnotemark \footnotetext{caused by blending of the $\textrm{[O\,\sc iii}]$${\lambda}$4364 emission line with the red wing of the broad H${\gamma}$ emission line. See D11 for details.} 
(See Section 2.2 and Figure 3). Collectively, the evidence presented supporting a centrally located H${^+}$ region in NGC 3998 does seem rather compelling and may be worth exploring for other Liners and Seyferts, especially those for which high angular resolution spectra exist. The astrophysics of 
H${^+}$ regions is well understood \citep{Osterbrock1989} and H${^+}$ regions are a widely observed phenomenon. 

Another important difference between LOC and XSTAR models is the geometry employed; slabs versus spheres, or open versus closed geometry. Open geometry predicts emission line flux ratios whereas closed geometry implies a covering factor yielding an important additional constraint, luminosity. Solutions to all photoionization models are highly degenerate. That is to say, for a given metallicity, there are many ${\rho_o, n}$ combinations that give rise to virtually identical emission line ratios and vice versa. However, a temperature and metallicity insensitive luminosity type parameter such as the extinction corrected H Balmer line luminosity is one important way to break the degeneracy, favouring photoionization models that employ closed geometry.

Finally, the gas pressure gradient in the H${^+}$ region of the best fitting model is a factor of ${\sim}$ 10${^4}$ smaller than required for hydrostatic equilibrium, thus an inflow at the free-fall velocity is inevitable.
The term ${outflow}$ is in widespread use in the professional literature, often with little or no qualification. A true outflow would require gas to exceed the escape velocity. Thus, any outflow velocity that is less than the escape velocity will inevitably become an inflow.

\subsection{Chaotic Cold Accretion}

Assuming spherical symmetry, the BLR of NGC 3998 contains ${\sim}$36 ${\times}$ 10$^3$ M${_{\sun}}$ of 
ionized H\footnotemark \footnotetext{superseding the lower value estimated by D11}, estimated by integrating the radial distribution, illustrated in Figure 5, over the radii 
2 ${\times}$ 10$^{-2}$ ${\le}$ r(pc) ${\le}$ 7.
How such a large mass of dust-free low metallicity gas should 
find its way into the centre of NGC 3998 can be understood in the context of chaotic cold accretion (CCA)
described by \cite{Gaspari2013} and \cite{Gaspari2015}. Briefly, CCA explains how 10$^4$ K gas can condense out of a 10$^6$ K circumgalactic halo and fall towards the central supermassive BH. 
In the absence of strong turbulence the inflowing gas is essentially in free-fall. In general, CCA can explain the low duty-cycle observed for AGN and, more specifically for NGC 3998, redirection of the jets powering the larger scale radio lobes \citep{Frank2016a}.

Assuming steady-state accretion, a mass inflow rate, $\dot{m}$ can be estimated using the equation of continuity
\begin{equation}
\dot{m(r)}  =  \epsilon 4 \pi r^2 v(r)  n_p(r) m_p
\end{equation}
The best fitting model employed a unity filling factor ${\epsilon}$ = 1, an inner radius, ${r}$ = 2 ${\times}$ 10$^{-2}$ pc, where the escape velocity, $v$ is determined by the mass distribution to be 10,950 km s$^{-1}$. The proton mass $m_p$ = 1.67 ${\times}$ 10$^{-27}$ kg. The proton number density ${n_p}$       
= 1 ${\times}$10$^5$ cm$^{-3}$ at the inner radius, leading to $\dot{m}$ = 1.4 ${\times}$ 10$^{-2}$ M${_{\sun}}$/yr, a factor
of 4.6 lower than a previous estimate in D11. The revised mass inflow rate is 
a factor of ${\sim}$2 larger than estimated by \cite{Nemmen2014}, a factor of 70 larger than
estimated by \cite{Yu2011}, and a factor of ${\sim}$4 larger than 
that required to explain the observed X-ray emission in terms of metrics
provided by \citet[see D11 for details]{Merloni2003}. The ADAF is located inside the Balmer emitting region, ${r}$ ${\leq}$ 2 ${\times}$ 10$^{-2}$ pc, consistent with the transition radius $R_{tr}$ estimated by \cite{Yu2011} but much smaller
than the one adopted by \cite{Nemmen2014}. 
Collectively, the model 
refines the initial conditions $\dot{m}$, $R_{tr}$ as well as the physical conditions of the inflowing gas powering the ADAF \citep{Narayan1994}. Interestingly, the mass inflow rates calculated
above are much higher than estimates for the rate of mass loss produced by the jets \citep{Nemmen2014, Yu2011}. What becomes of the excess? Hopefully, this question will be answered by examining jet models
\citep[e.g.][]{Tchekhovskoy2016}.

\section{Conclusions}

A new physical model describes the 
H${\alpha}$ emission line profile shape, as well as the luminosities and relative intensities of the H Balmer, [O\,{\sc iii}], and quite likely several of the broad UV emission lines seen in the STIS spectra of the LINER 1.2 nucleus of NGC 3998, in terms of an H${^+}$ region consisting of a large volume of low density gas of high ionization parameter that is intrinsically dust-free due to low metallicity. The H${\alpha}$ emission line profile shape is consistent with only the BH mass
measured by \citet{Francesco2006} because the larger BH mass of \citet{Walsh2012a} produced model H${\alpha}$ emission line profiles that consistently did not match the observed one.
The photoionization
calculations indicate that the BLR, synonymous now with the H${^+}$ region, is large, ${\sim}$7 pc in radius, identical to a prior estimate by D11,
causing NGC 3998 to not conform to the BLR size-luminosity correlation established for reverberating AGN.  The BLR of NGC 3998 contains ${\sim}$36 ${\times}$ 10$^3$ M${_{\sun}}$ of ionized gas. How such a large mass of dust-free low metallicity gas should 
find its way into the centre of NGC 3998 can be understood in the context of chaotic cold accretion.
Collectively, the model refines values of the initial conditions
needed for further modelling of the ADAF in NGC 3998, that, in turn, inform jet models. 
\section*{Acknowledgements}

The author thanks Prof. R. Nemmen for the ADAF spectrum illustrated in Figure 1, Dr. Marc Sarzi for
providing the stellar mass model presented in Table 2, Dr. Tim Kallman for help with XSTAR, Dr. Steven Willner for patiently reviewing and commenting on many drafts of the manuscript, and lastly the referee for thought-provoking questions and comments.




\bibliographystyle{mnras}
\bibliography{N3998} 






\bsp	
\label{lastpage}
\end{document}